# Labeling Workflow Views with Fine-Grained Dependencies

Zhuowei Bao
Department of Computer and
Information Science
University of Pennsylvania
Philadelphia, PA 19104, USA
zhuowei@cis.upenn.edu

Susan B. Davidson
Department of Computer and
Information Science
University of Pennsylvania
Philadelphia, PA 19104, USA
susan@cis.upenn.edu

Tova Milo
School of Computer Science
Tel Aviv University
Tel Aviv, Israel
milo@cs.tau.ac.il

## ABSTRACT

This paper considers the problem of efficiently answering reachability queries over *views* of provenance graphs, derived from executions of workflows that may include recursion. Such views include composite modules and model fine-grained dependencies between module inputs and outputs. A novel *view-adaptive* dynamic labeling scheme is developed for efficient query evaluation, in which view specifications are labeled statically (i.e. as they are created) and data items are labeled dynamically as they are produced during a workflow execution. Although the combination of fine-grained dependencies and recursive workflows entail, in general, long (linear-size) data labels, we show that for a large natural class of workflows and views, labels are *compact* (logarithmic-size) and reachability queries can be evaluated in constant time. Experimental results demonstrate the benefit of this approach over the state-of-the-art technique when applied for labeling multiple views.

## 1. INTRODUCTION

The ability to manage workflow provenance is increasingly important for scientific as well as business applications. For example, if an input to a workflow execution is discovered to be incorrect, we may wish to determine whether a particular workflow output depends on it and is thus also potentially incorrect. Finding efficient techniques to answer such *reachability* queries is thus of particular interest.

However, provenance information can be extremely large, so we may wish to provide different *views* of this information. For example, users may wish to specify *abstraction views* which focus user attention on relevant provenance information and abstract away irrelevant details, an idea proposed in [8]. Workflow owners may also wish to specify *security views* which can be used to hide private information from certain user groups (e.g., sensitive intermediate data and module functionality [10]). Provenance views consist of a set of *composite modules* which encapsulate subworkflows.

EXAMPLE 1. *Figure 1 shows an abstraction of a real-life scientific workflow collected from the myExperiment reposi-*

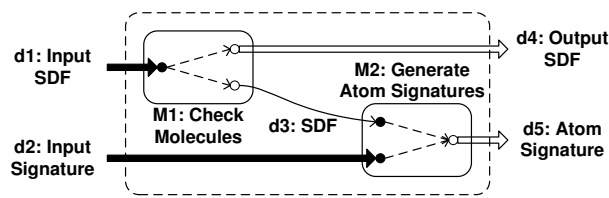

**Figure 1: Views with Fine-Grained Dependencies**

*tory [19]. It generates atom signatures for individual compounds given a Structural Data File (SDF) as input (ignore for now the dashed edges inside modules $M_1$ and $M_2$). In a high-level view of this workflow, users see only one composite module, indicated as the big dashed box, with two inputs ($d_1$ and $d_2$) and two outputs ($d_4$ and $d_5$), while modules $M_1$ and $M_2$ and intermediate data $d_3$ are hidden.*

An important thing to keep in mind is that Workflow provenance not only records the order of module executions but also the dependencies between inputs and outputs of modules. Therefore, workflow views should explicitly specify the input-output dependencies for modules that are exposed to users. Previous research [13, 21, 4, 5] has adopted a simplified provenance model which assumes that every output of a module depends on every input, termed *black-box dependencies*. However, a more fine-grained provenance model captures the fact that the output of a module may depend on only a subset of its inputs.

To understand why fine-grained dependencies are useful, consider the two types of views mentioned earlier. In abstraction views, although irrelevant workflow details are hidden inside composite modules, users should still be able to see the true dependencies between inputs and outputs of composite modules (*white-box dependences*). In security views, however, one may want to hide the true dependencies between inputs and outputs of certain composite modules in order to preserve structural or module privacy [10]. To this end, one may move to somewhere on the spectrum between white-box and black-box dependencies (*grey-box dependencies*). With grey-box dependencies, additional (false) dependencies between inputs and outputs may be added.

EXAMPLE 2. *Returning to Figure 1, fine-grained dependencies between the inputs and outputs of modules $M_1$ and $M_2$ are indicated as dashed edges inside the modules. In an abstraction view, the composite module would be associated with white-box dependencies, in which $d_4$ depends on $d_1$ but not on $d_2$. However, in a security view, the composite module could be associated with a grey-box dependency matrix in which every output depends on every input. Hence, the answer to the reachability query "Does $d_4$ depend on $d_2$?" is different in the two views.*

This paper considers the problem of efficiently answering reachability queries over views of provenance graphs, of the types illustrated above. A common approach for processing reachability queries is to label data items so that the reachability between any two items can be answered efficiently by comparing their labels. Moreover, data items must be labeled dynamically as soon as they are produced during the execution, since scientific workflows can take a long time to execute and users may wish to query partial executions.

In contrast to previous work, we study effective dynamic labeling in the context of (1) *fine-grained dependencies* between inputs and outputs of modules; and (2) *views with grey-box dependencies*. This context introduces several new challenges. First, none of the existing dynamic labeling schemes applies to fine-grained dependencies, since they all rely on a simplified provenance model with black-box dependencies. Second, due to grey-box dependencies, the answer to a reachability query may alter in different views. A brute-force approach to handling multiple views is to label data items for each view repeatedly and separately. This has two drawbacks: (i) large index: for each data item, we must maintain one label for each view; and (ii) expensive index maintenance: when a new view is added, all existing data items must be re-labeled. To address the challenges, more effective labeling techniques must be developed. The main contributions of this paper are summarized as follows.

• We propose a formal model based on graph grammars which capture a rich class of (possibly recursive) workflows with fine-grained dependencies between the inputs and outputs of modules. We then use the model to formalize the notion of views. They are defined over the workflow specification and then naturally projected onto its runs (Section 2).

• To get a handle on the difficulty introduced by fine-grained dependencies to the dynamic labeling problem, we prove that in general, long (linear-size) labels are required. We further show that common restrictions on the workflow specification, that sufficed to reduce the label length for black-box dependencies [5], are no longer helpful. Nevertheless, we identify a large natural class of *safe views* over *strictly linear-recursive* workflows for which dynamic, yet compact (logarithmic-size) labeling is possible (Section 3).

• Based on this foundation we propose a novel labeling approach whereby view specifications are labeled *statically* (i.e. as they are created), whereas data items are labeled *dynamically* as they are produced during a workflow execution. At query time, the labeling of the view over which the reachability query is asked is used to augment the data labels to provide the correct answer in constant time. We call this a *view-adaptive* dynamic labeling scheme. It has the great advantage that, since data labels are unrelated to any view, views can be added/deleted/modified without having to touch the data. It is both space-efficient and time-efficient relative to the brute-force approach (Section 4).

• Finally, we evaluate the proposed view-adaptive labeling scheme over both real-life and synthetic workflows. The experimental study demonstrates the superiority of our view-adaptive labeling approach over the state-of-the-art technique [5] when applied to label multiple views (Section 5).

**Related Work.** Before presenting our results, we briefly review related work. The problem of reachability labeling has been studied for different classes of graphs in both static and dynamic settings. Ideally, one would like to build *compact* (logarithmic-size) labels which enable *efficient* (constant) query processing. While compact and efficient labeling is shown to be feasible for static trees [20], when labeling general directed acyclc graphs (DAGs), any possible scheme requires linear-size labels even if arbitrary query time is allowed [4]. On the other hand, dynamic labeling is also much harder than static labeling. [9] shows that even labeling dynamic trees requires linear-size labels. Fortunately, although workflow runs can have arbitrarily more complex DAG structures than trees, [4, 5] show that knowledge of the specification can be exploited to obtain compact and efficient labeling schemes for both static and dynamic runs derived from a given specification. A more detailed comparison between existing static and dynamic labeling schemes for XML trees [20, 1, 9, 18, 23], for DAGs [15, 24, 22, 16, 11] and for workflow runs [13, 4, 5] is summarized in [5]. However, as mentioned above, none of the existing dynamic labeling schemes is applicable to our problem as they neither support fine-grained dependencies nor handle views.

## 2. MODEL AND PROBLEM STATEMENT

We present a fine-grained workflow model with white-box dependencies in Section 2.1. Based on this model, we define views with grey-box dependencies in Section 2.2. Section 2.3 formulates the view-adaptive dynamic labeling problem.

### 2.1 Fine-Grained Workflow Model

Our workflow model is built upon two concepts: *workflow specification*, which describes the design of a workflow, and *workflow run*, which describes a particular workflow execution. We model the structure of a specification as a *context-free workflow grammar* whose *language* corresponds to exactly the set of all possible runs of this specification. The grammar that we use is similar to [5, 7]. However, previous work [17, 13, 21, 5, 7] adopted a simplified provenance model which implicitly assumes *black-box dependencies* – every output of a module depends on every input. In contrast, this paper proposes a more *fine-grained* provenance model which captures the fact that an output of a module may depend on only a subset of inputs. We call this *white-box dependencies*. In particular, our model associates the grammar with a *dependency assignment* that explicitly specifies the dependencies between inputs and outputs of atomic modules.

The basic building blocks of our model are *modules* and *simple workflows*. A module has a set of input ports and a set of output ports; and a simple workflow is built up from a set of modules by connecting their input and output ports.

*Definition 1.* (**Module**) A *module* is $M = (I, O)$, where $I$ is a set of *input ports* and $O$ is a set of *output ports*.

*Definition 2.* (**Simple Workflow**) A *simple workflow* is $W = (V, E)$, where $V$ is a multiset of *modules* and $E$ is a set of *data edges* from an output port of one module to an input port of another module. Each data edge carries a unique *data item* that is produced by the former and then consumed by the latter. Input ports with no incoming data edges are called *initial input ports*; and output ports with no outgoing data edges are called *final output ports*.

To simplify the presentation, we assume that (1) *pairwise non-adjacent data edges*: any pair of data edges are not incident to the same port; and (2) *acyclic simple workflow*: data edges do not form cycles among the modules. Note that the above two restrictions do not limit the expressive power of our model. For (1), adjacent data edges can be resolved by introducing dummy modules that distribute or aggregate multiple data items. For (2), we will see that loops can be implicitly captured by recursive productions.

EXAMPLE 3. *The top left corner of Figure 2 shows a module $S$ with two input ports and three output ports, which are denoted by solid and empty cycles, respectively. The top right corner of Figure 2 shows a simple workflow $W_1$ with six modules and ten data edges (solid edges, ignore the dashed edges inside modules for now). $W_1$ has two initial input ports and three final output ports, which are highlighted by solid and empty thick arrows, respectively.*

To build a new workflow, an existing (simple) workflow may be reused as a *composite module*. This is modeled by a *workflow production*.

*Definition 3.* **(Workflow Production)** A *workflow production* is of form $M \to_f W$, where $M$ is a composite module, $W$ is a simple workflow and $f$ is a bijection that maps input ports and output ports of $M$ to initial input ports and final output ports of $W$, respectively. When $f$ is clear from the context, we simply denote a production by $M \to W$.

EXAMPLE 4. *In Figure 2, each row defines one or two productions. For example, the first row defines $S \to W_1$, and the second row defines $A \to W_2$ and $A \to W_3$. Note that $A$ also appears as a composite module in both $W_1$ and $W_4$. For simplicity, we assume that for each production $M \to W$, the (initial) input ports and (final) output ports of $M$ and $W$ are mapped by $f$ from top to bottom as shown in the figure.*

The *context-free workflow grammar* is a natural extension of the well-known context-free string grammar, where modules correspond to characters, and simple workflows that are built up from modules correspond to strings that are sequences of characters. In particular, atomic and composite modules correspond to terminals and variables, respectively. We also define a start module and a finite set of workflow productions. By Definition 3, each production $M \to_f W$ replaces a composite module $M$ with a simple workflow $W$. The data edges adjacent to $M$ are connected to $W$ based on the bijection $f$. The *language* of a context-free workflow grammar consists of all simple workflows that can be derived from the start module and contain only atomic modules.

Following the standard notations for string grammars, given a finite set $\Sigma$ of modules, let $\Sigma^*$ denote the set of all simple workflows that are built up from a multiset of modules in $\Sigma$. Given two simple workflows $W_1$ and $W_2$, let $W_1 \Rightarrow_f^* W_2$ denote that $W_2$ can be derived from $W_1$ by applying a sequence of zero or more productions, and $f$ is a bijection that maps initial input ports and final output ports from $W_1$ to $W_2$. Again, $f$ may be omitted for simplicity.

*Definition 4.* **(Context-Free Workflow Grammar)** A *context-free workflow grammar* (abbr. *workflow grammar*) is $G = (\Sigma, \Delta, S, P)$, where $\Sigma$ is a finite set of modules, $\Delta \subseteq \Sigma$ is a set of *composite modules* (then $\Sigma \setminus \Delta$ is the set of *atomic modules*), $S \in \Sigma$ is a *start module*, and $P = \{M \to W \mid M \in \Delta, W \in \Sigma^*\}$ is a finite set of *workflow productions*. The *language* of $G$ is $L(G) = \{R \in (\Sigma \setminus \Delta)^* \mid S \Rightarrow^* R\}$.

EXAMPLE 5. *Our running example of a workflow grammar $G$ is shown in Figure 2. Composite modules are indicated by uppercase letters and atomic modules by lowercase letters. Formally, $G = (\Sigma, \Delta, S, P)$, where $\Sigma = \{S, A, B, \ldots, E, a, b, \ldots, f\}$, $\Delta = \{S, A, B, \ldots, E\}$, and $P = \{p_1 = S \to W_1,\ p_2 = A \to W_2,\ p_3 = A \to W_3,\ p_4 = B \to W_4,\ p_5 = C \to W_5,\ p_6 = D \to W_6,\ p_7 = D \to W_7,\ p_8 = E \to W_8\}$. Note that $p_2$ and $p_4$ form a recursion between $A$ and $B$. $p_6$ forms a self-recursion over $D$, and along with $p_7$, indicates a loop (sequential execution) over $f$.*

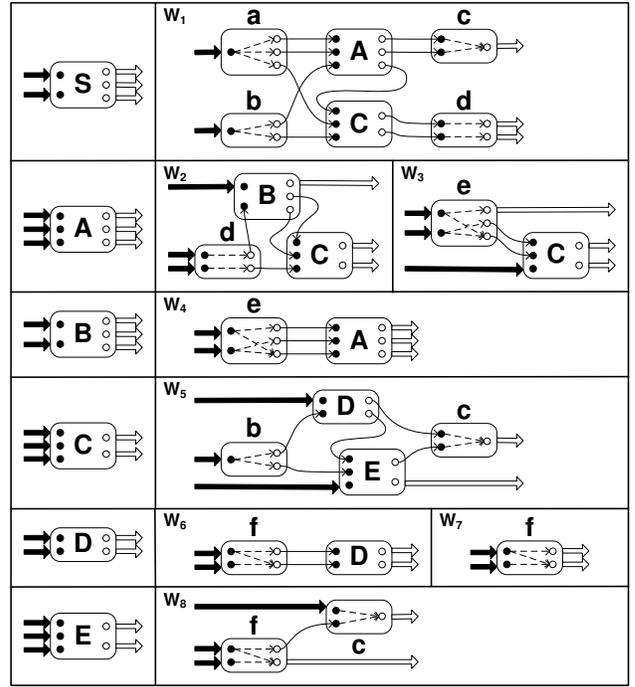

Figure 2: **Workflow Specification**

*One possible simple workflow run $R \in L(G)$ is shown in Figure 3, where the atomic modules in $R$ are denoted by solid boxes, and the composite modules that are created during the derivation of $R$ are denoted by dashed boxes. We create a unique id for each atomic and composite module in $R$ by appending a distinct number to the module name. $d_1, d_2, \ldots, d_{41}$ are unique ids for data items (data edges) in $R$. For sake of illustration, we omit details of $C\!:\!1$, $C\!:\!2$ and $C\!:\!3$, and show details of $C\!:\!4$ in Figure 4. Observe that $R$ can be derived from $S$ by applying a sequence of productions $p_1, p_2, p_4, p_2, p_4, p_3, p_5, p_6, p_6, p_7, p_8, \ldots$*

So far we consider only workflow structure – the way in which modules are connected to construct workflows. Next, we enrich the model by defining fine-grained dependencies between inputs and outputs of atomic modules. Naturally, we assume that every input contributes to at least one output; and every output depends on at least one input.

*Definition 5.* **(Dependency Assignment)** Given a finite set $\Sigma$ of modules, a *dependency assignment* to $\Sigma$ is a function $\lambda$ that, for each module $M = (I, O) \in \Sigma$, defines a set $\lambda(M)$ of *dependency edges* from $I$ to $O$, such that $\forall i \in I, \exists o \in O, (i, o) \in \lambda(M)$; and $\forall o \in O, \exists i \in I, (i, o) \in \lambda(M)$.

Finally, combining all the above components, our fine-grained workflow model is formalized as follows.

*Definition 6.* **(Fine-Grained Workflow Model)** A *workflow specification* is $G^\lambda$, where $G = (\Sigma, \Delta, S, P)$ is a workflow grammar and $\lambda$ is a dependency assignment to $\Sigma \setminus \Delta$. The set of all *workflow runs* w.r.t. $G^\lambda$ is $L(G^\lambda) = \{R^\lambda \mid R \in L(G)\}$, where $R^\lambda$ is obtained from $R$ by adding to each module $M$ in $R$ a set $\lambda(M)$ of dependency edges.

EXAMPLE 6. *For the grammar $G$ in Figure 2, we define a dependency assignment $\lambda$ to all atomic modules (i.e., $a$, $b$, …, $f$). The dependency edges introduced by $\lambda$ are shown in Figure 2 as dashed edges from input ports to output ports of atomic modules. With both data (solid) and dependency (dashed) edges, Figures 3 and 4 represent a run $R^\lambda \in L(G^\lambda)$.*

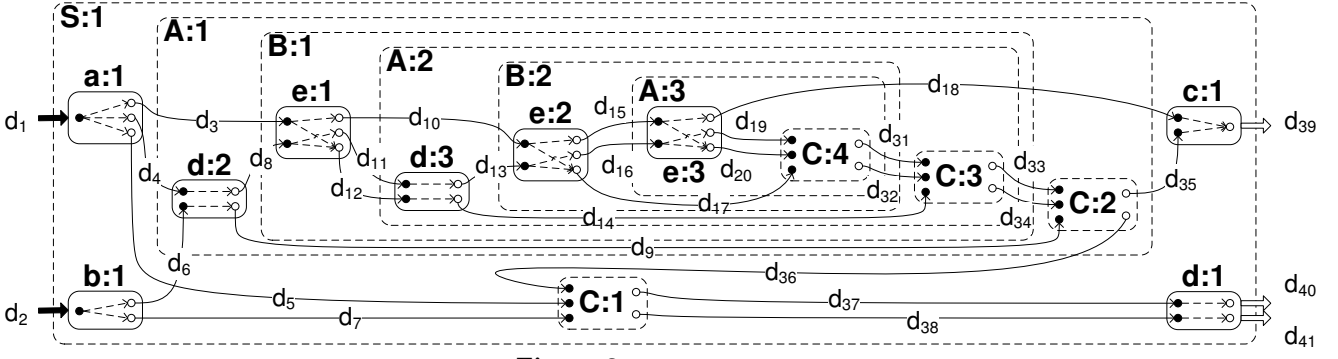

Figure 3: Workflow Run

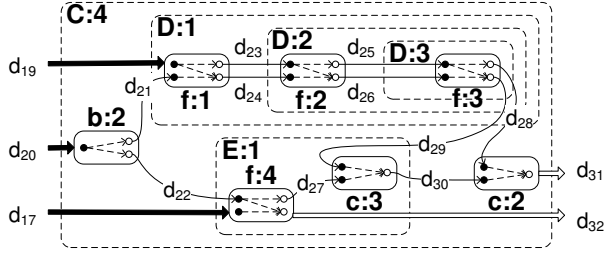

Figure 4: Details of Composite Module C:4

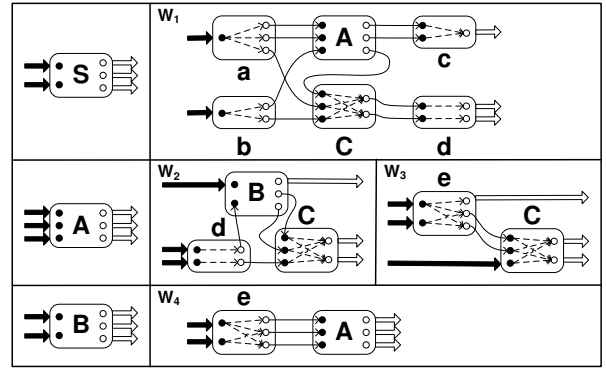

Figure 5: View of Workflow Specification

In Section 3, we will compare our fine-grained model (i.e., with white-box dependencies) to the existing coarse-grained model (i.e., with black-box dependencies) [5, 7]. Both are grammar-based, but the coarse-grained model is less expressive, and captures only a subclass of fine-grained workflows.

*Definition 7.* (**Coarse-Grained Workflows**) A workflow specification $G^\lambda$ is said to be *coarse-grained* if (1) $\lambda$ is defined such that for any atomic module, every output depends on every input; and (2) every simple workflow used by $G$ has a single source module and a single sink module [1].

## 2.2 Views with Grey-Box Dependencies

A *workflow view* is constructed over a specification and then projected onto its runs. Such approach is common in workflows [8, 21, 10] (unlike typical database views that are defined via queries), but our work is the first to be based on a fine-grained model. Formally, a view is defined by two components. One describes the structure of a view by restricting the possible expansions of workflow hierarchy to a subset of composite modules. The other specifies the "perceived" fine-grained dependencies between inputs and outputs of all unexpandable modules in this view. As mentioned in Section 1, for *abstraction views*, the perceived dependencies always reflect the true dependencies, which we call *white-box dependencies*. In contrast, for *security views*, false dependencies may be introduced in order to hide private provenance information, which we call *grey-box dependencies*.

*Definition 8.* (**Workflow View**) Give a workflow specification $G^\lambda = (\Sigma, \Delta, S, P)^\lambda$, a *view* over $G^\lambda$ is defined by a pair $(\Delta', \lambda')$, where $\Delta' \subseteq \Delta$ is a subset of composite modules and $\lambda'$ is a new dependency assignment for $\Sigma \setminus \Delta'$. In particular, $(\Delta, \lambda)$ is said to be the *default view* over $G^\lambda$.

REMARK 1. *As will be seen in Section 3.1, from the input-output dependencies of atomic modules, we can compute those of composite modules. We thus say that a view $(\Delta', \lambda')$ has* white-box dependencies, *if $\lambda'$ defines the same dependencies as $\lambda$ does, otherwise, it has* grey-box dependencies.

[1](2) ensures black-box dependencies for composite modules.

A view $U = (\Delta', \lambda')$ defined over a specification $G^\lambda$ produces a new grammar, denoted $G_{\Delta'}$, by restricting $G$ to the subset of productions for composite modules in $\Delta'$. Together with $\lambda'$, it defines a new specification, denoted $G_U = (G_{\Delta'})^{\lambda'}$, which we call a *view of this specification*. Similarly, given a run $R^\lambda \in L(G^\lambda)$, by restricting the derivation of $R$ to only productions for composite modules in $\Delta'$ and using $\lambda'$, we obtain a *view of this run*, denoted $R_U = (R_{\Delta'})^{\lambda'}$.

EXAMPLE 7. *Using the specification $G^\lambda$ in Figure 2, we define a view $U = (\Delta', \lambda')$, where $\Delta' = \{S, A, B\}$. The new grammar $G_{\Delta'}$ is shown in Figure 5, which contains only the productions for $S$, $A$ and $B$. Note that $C$ is treated as an atomic module in this view, which makes $D$, $E$ and $f$ underivable. Therefore, $\lambda'$ needs to be defined for only atomic modules $a$, $b$, $c$, $d$, $e$ and $C$. The dependency edges introduced by $\lambda'$ are shown in Figure 5 as dashed edges. Comparing with $\lambda$ defined in Figure 2, we observe that $\lambda'(C)$ is newly defined, $\lambda'(e)$ is changed, and others are unchanged. Hence, this view introduces grey-box dependencies.*

*We project this view onto the run $R^\lambda$ in Figures 3 and 4. Since $C$ is treated as atomic, details of $C:1$, $C:2$, $C:3$ and $C:4$ (Figure 4) are hidden and $R_{\Delta'}$ has exactly the structure in Figure 3. However, all the dependency edges for $R_{\Delta'}$ should be given according to $\lambda'$ as in Figure 5.*

In the rest of this paper, we may simply denote a specification by $G$ and a run by $R$, since the original dependency assignment $\lambda$ is irrelevant to views (i.e., overwritten by $\lambda'$).

## 2.3 View-Adaptive Dynamic Labeling

We start with the basic dynamic labeling problem. The goal is to assign each data item a *reachability label* as soon as it is produced (*dynamically*) such that using only the labels of any two data items, we can quickly decide if one depends on the other. Two different but related dynamic labeling problems were formulated in [5]. In the *execution-based* problem, atomic modules of a run are generated one-by-one according to some topological ordering. In the *derivation-based* problem, a run is derived from the start module by applying a sequence of productions. As observed in [5], any solution for the former also provides a solution for the latter. We thus focus only on the derivation-based problem.

*Definition 9.* [5] **(Dynamic Labeling)** A *dynamic labeling scheme* for a given specification $G^\lambda$ is $(\phi, \pi)$, where $\phi$ is a labeling function and $\pi$ is a binary predicate. $\phi$ takes as input a derivation of a run $R^\lambda \in L(G^\lambda)$, that is, a sequence of productions that transform the start module $S$ to $R$. Initially, $\phi$ assigns a label $\phi(d)$ to each input and output $d$ of $S$. In the $i$th step of the derivation, $\phi$ assigns a label $\phi(d)$ to each new data item $d$ introduced by the $i$th production. Note that we do not know the production sequence in advance, but receive them online. The assigned labels cannot be modified subsequently. $\phi$ and $\pi$ are such that for any derivation of a run $R^\lambda \in L(G^\lambda)$ and any two data items $d_1$ and $d_2$ in $R^\lambda$, $\pi(\phi(d_1), \phi(d_2)) = \texttt{true}$ iff $d_2$ depends on $d_1$.

In contrast to the previous work [5], this paper studies the dynamic labeling problem in more general and useful workflow settings. Specifically, we consider (1) *fine-grained input-output dependences* and (2) *views with grey-box dependencies*. Both ingredients entail new challenges, which will be addressed in Sections 3 and 4, respectively.

To handle views, we propose in Section 4 a novel *view-adaptive* labeling approach whereby view specifications are labeled *statically* (i.e., as they are created), whereas data items are labeled *dynamically* as they are produced during a workflow execution. At query time, the label of the view over which the query is asked is combined with the labels of relevant data items to provide the correct answer. In this framework, since data labels are unrelated to any view (*view-adaptive*), views can be added/deleted/modified without having to touch the data. It is both space-efficient and time-efficient relative to the alternative approach where data items are labeled repeatedly and separately for each view.

*Definition 10.* **(View-Adaptive Dynamic Labeling)** A *view-adaptive* dynamic labeling scheme for a given specification $G$ is $(\phi_r, \phi_v, \pi)$, where $\phi_r$ is a labeling function for runs, $\phi_v$ is a labeling function for view specifications, and $\pi$ is a ternary predicate. Given a derivation of a run $R \in L(G)$, $\phi_r$ as before assigns a label $\phi_r(d)$ (called *data label*) to each data item $d$ as soon as it is produced during the derivation of $R$. Given a view $U$ over $G$, $\phi_v$ treats $U$ as one object and assigns a label $\phi_v(U)$ (called *view label*). $\phi_r$, $\phi_v$ and $\pi$ are such that for any derivation of a run $R \in L(G)$, any view $U$ over $G$ and any two data items $d_1$ and $d_2$ in $R_U$, $\pi(\phi_r(d_1), \phi_r(d_2), \phi_v(U)) = \texttt{true}$ iff $d_2$ depends on $d_1$ w.r.t. $U$.

A (view-adaptive) dynamic labeling scheme is said to be *compact* if for any derivation of a run with $n$ data items, it creates data labels of $O(\log n)$ bits. Clearly, it provides shortest possible data labels up to a constant factor.

## 3. FEASIBILITY OF DYNAMIC LABELING

To address the challenges brought by fine-grained dependencies, we first consider the basic dynamic labeling problem (see Definition 9), where there is only one default view defined over the specification. Note that the labels created for the default view also work for other views with white-box dependencies, but not those with grey-box dependencies.

As a formal analysis, we present in this section a classification of fine-grained workflows based on the feasibility of developing (compact) dynamic labeling schemes. In Section 3.1, we first identify a class of *safe* workflows, and show that they are the largest set of workflows that allow dynamic labeling schemes. In Section 3.2, we further identify a class of *strictly linear-recursive* workflow structures for which dynamic, yet *compact* labeling schemes are possible. Polynomial-time algorithms are also given to decide if a workflow is safe or if its structure is strictly linear-recursive.

Interestingly, our results show that the common restriction on the workflow structure, which sufficed to reduce the label length for black-box dependencies [5], are no longer helpful. This formally proves the difficulty introduced by fine-grained dependencies to the dynamic labeling problem.

### 3.1 Safe Workflows

Some workflows cannot be labeled on-the-fly even if arbitrary label size is allowed. We illustrate by an example.

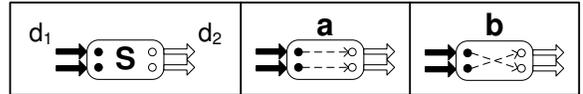

**Figure 6: Unsafe Workflow**

EXAMPLE 8. *Consider the specification in Figure 6 with two productions $S \to a$ and $S \to b$. $d_1$ and $d_2$ are an input and an output of $S$, respectively. Observe that if $S \to a$ is applied, then $d_2$ depends on $d_1$; otherwise (if $S \to b$ is applied), $d_2$ does not depend on $d_1$. Recall from Definition 9 that the labels for $d_1$ and $d_2$ must be assigned before we see the production, and cannot be modified subsequently. Therefore, no dynamic labeling schemes exist for this example.*

In general, if two simple workflows with only atomic modules can be derived from the same composite module, and they are *inconsistent*, in the sense that they have different dependencies between initial inputs and final outputs, then dynamic labeling is impossible for this specification. Such workflows are said to be *unsafe*, and the others are *safe*.

*Definition 11.* **(Safe Workflow)** A workflow specification $G^\lambda = (\Sigma, \Delta, S, P)^\lambda$ is said to be *safe* if $\forall M \in \Delta$ and $W_1, W_2 \in (\Sigma \setminus \Delta)^*$ such that $M \Rightarrow^* W_1$ and $M \Rightarrow^* W_2$, $W_1$ is consistent with $W_2$ w.r.t. $\lambda$. Also, $\lambda$ is said to be *safe* if $G^\lambda$ is safe; and a view $U$ is said to be *safe* if $G_U$ is safe.

REMARK 2. *Safety is a natural restriction on fine-grained workflows. It essentially says that for any module, either atomic or composite, the dependences between inputs and outputs are deterministic, in the sense that they can be predicted from the specification, and are consistent among all possible executions. In particular, by Definition 7, any coarse-grained workflow (i.e., with black-box dependencies) is always safe. Moreover, it is important to notice that from the perspective of data provenance, the output of an aggregate function depends on each of its inputs [3], even though the output may take the value from only one of its inputs (e.g., "max" or "min" functions). Therefore, a workflow that use those aggregate functions as modules is still safe.*

Our first result shows that safety characterizes the feasibility of dynamic labeling for fine-grained workflows.

THEOREM 1. *Given any workflow specification $G^\lambda$, there is a dynamic labeling scheme for $G^\lambda$ iff $G^\lambda$ is safe.*

PROOF. (Sketch) By Definition 11, unsafe workflows do not allow any dynamic labeling schemes. On the other hand, the view-adaptive dynamic labeling scheme, which we will present in Section 4, can be modified to label arbitrary safe workflows, though it may create linear-size data labels. □

It is possible to test in polynomial time if a given specification $G^\lambda$ is safe. Our algorithm based on Lemma 1 is briefly described as follows. We start by defining $\lambda^* = \lambda$ for each atomic module, and then compute $\lambda^*$ for composite modules by verifying all the productions. A production $M \to W$ is said to be *verifiable*, if $\lambda^*$ is already defined for all the modules in $W$, so that $\lambda^*(M)$ can be computed. The algorithm reports that $G^\lambda$ is safe, if $\lambda^*$ is consistently defined for all composite modules, and outputs $\lambda^*$ as a by-product.

LEMMA 1. **(Full Assignment)** *A workflow specification $G^\lambda = (\Sigma, \Delta, S, P)^\lambda$ is safe iff there is a unique dependency assignment $\lambda^*$ to $\Sigma$ (called the* full dependency assignment*) such that (1) $\forall M \in \Sigma \setminus \Delta$, $\lambda^*(M) = \lambda(M)$; and (2) $\forall M \to W \in P$, $M$ is consistent with $W$ w.r.t. $\lambda^*$.*

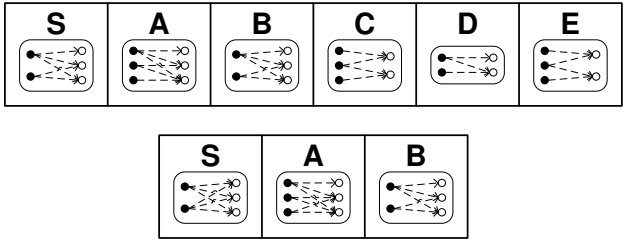

**Figure 7: Full Dependency Assignment**

EXAMPLE 9. *We illustrate the above algorithm using the specification $G^\lambda$ in Figure 2. Initially, both $p_7 = D \to W_7$ and $p_8 = E \to W_8$ are verifiable. We compute $\lambda^*(D)$ and $\lambda^*(E)$ by $p_7$ and $p_8$. Once $\lambda^*(D)$ and $\lambda^*(E)$ are defined, $p_5 = C \to W_5$ and $p_6 = D \to W_6$ become verifiable. We compute $\lambda^*(C)$ by $p_6$, and verify that $\lambda^*(D)$ computed by $p_6$ is consistent with the one computed before by $p_7$. We continue this process until all the productions are verified. Hence, $G^\lambda$ is safe, and $\lambda^*$ is shown on the top of Figure 7. Similarly, one can verify that the view $U = (\Delta', \lambda')$ defined in Example 7 is safe using Figure 5. The full dependency assignment for $U$ is shown on the bottom of Figure 7. Comparing the two full assignments in Figure 7, while $B$ gets the same dependencies, the ones for $S$ and $A$ are different.*

## 3.2 Linear-Recursive Workflow Structures

For safe workflows, we further examine the feasibility of developing compact dynamic labeling schemes. First of all, a negative result in [5] shows that there is a coarse-grained workflow that does not allow any compact dynamic labeling scheme. By Definition 7 and Lemma 1, we know that any coarse-grained workflow is safe. So the negative result also applies to the fine-grained model: there is a safe workflow that does not allow any compact dynamic labeling scheme.

Given this, our next goal is to identify safe workflows that enable compact dynamic labeling. An elegant characterization for coarse-grained workflows is proved in [5]: given any coarse-grained workflow specification $G^\lambda$, there is a compact dynamic labeling scheme for $G^\lambda$ iff $G$ is a linear-recursive workflow grammar which is formally defined as follows.

*Definition 12.* [5] **(Linear-Recursive Workflow Grammar)** A workflow grammar $G = (\Sigma, \Delta, S, P)$ is said to be *linear-recursive* if $\forall M \in \Delta$ and $W \in \Sigma^*$ such that $M \Rightarrow^* W$, $W$ has at most one instance of $M$.

Note that coarse-grained workflows are only a restricted class of (fine-grained) safe workflows. We show here that, in the fine-grained model, linear-recursiveness is not enough to enable compact dynamic labeling for safe workflows.

THEOREM 2. *There is a linear-recursive grammar $G$ and a safe dependency assignment $\lambda$ such that any dynamic labeling scheme for $G^\lambda$ requires linear-size data labels.*

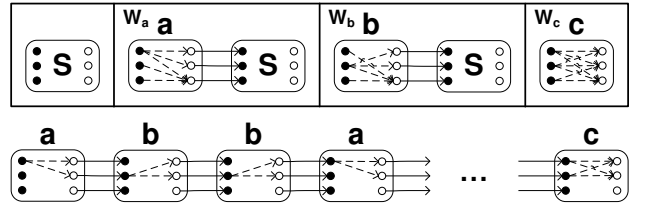

**Figure 8: Counterexample in Proof of Theorem 2**

PROOF. (Sketch) Figure 8 gives a counterexample $G^\lambda$ with three productions $p_a = S \to W_a$, $p_b = S \to W_b$ and $p_c = S \to W_c$, where $G$ is linear-recursive and $\lambda$ is safe. Observe that a run $R^\lambda \in L(G^\lambda)$ is derived from the start module $S$ by applying an arbitrary sequence of $p_a$ and $p_b$, followed by one $p_c$. Both $p_a$ and $p_b$ produce three new data items (data edges). We focus only on the dependency edges between the first two data items. Observe from Figure 8 that they form a binary tree that is created dynamically from left to right: if $p_a$ is applied, then the first data item is expanded, otherwise (if $p_b$ is applied), the second data item is expanded. Using a similar technique to [9], we can prove that labeling such a dynamic tree requires linear-size data labels. □

Theorem 2 tells us that while fine-grained dependencies increase the expressive power of the model, they limit the recursive workflow structure that allows compact dynamic labeling. We thus identify a natural class of *strictly linear-recursive* workflow grammars for which dynamic, yet compact labeling is feasible for any safe dependency assignment. To define them, we introduce a *production graph* that describes the derivation relationship between modules.

*Definition 13.* **(Production Graph)** Given a workflow grammar $G = (\Sigma, \Delta, S, P)$, the *production graph* of $G$ is a directed multigraph $\mathcal{P}(G)$ in which each vertex denotes a unique module in $\Sigma$. For each production $M \to W$ in $P$ and each module $M'$ in $W$, there is an edge from $M$ to $M'$ in $\mathcal{P}(G)$. Note that if $W$ has multiple instances of a module $M'$, then $\mathcal{P}(G)$ has multiple parallel edges from $M$ to $M'$.

Intuitively, every cycle in $\mathcal{P}(G)$ corresponds to a recursion in $G$. $G$ is said to be *recursive* if $\mathcal{P}(G)$ is cyclic. A module in $G$ is said to be *recursive*, if it belongs to a cycle in $\mathcal{P}(G)$.

*Definition 14.* **(Strictly Linear-Recursive Workflow Grammar)** A workflow grammar $G$ is said to be *strictly linear-recursive* if all the cycles in $\mathcal{P}(G)$ are vertex-disjoint.

REMARK 3. *Strictly linear recursion is able to capture common recursive patterns that we observed from the myExperiment workflow repository [19]. In particular, consider two common forms of recursion that we encounter in real-life scientific workflows. The first is called the* loop *execution for which a sub-workflow is repeated sequentially a number of times until certain condition is met. The second is called the* fork *execution for which multiple copies of a sub-workflow are executed in parallel. In scientific workflow systems, such as Taverna [14] and Kepler [2], fork executions are commonly used to model operations over complex data (e.g., "maps" over sets). Both loop and fork executions belong to a simple form of strictly linear recursion.*

It is easy to show that every strictly linear-recursive workflow grammar is also linear-recursive, but not vice versa.

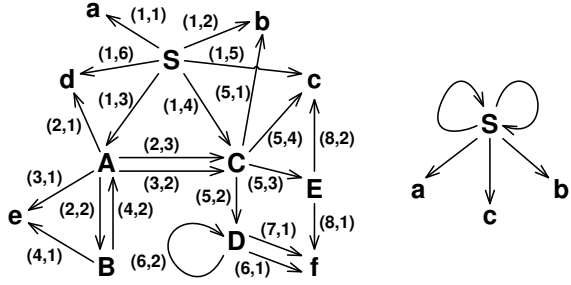

**Figure 9: Production Graphs**

EXAMPLE 10. *Figure 9 (left) shows the production graph $\mathcal{P}(G)$ for the grammar $G$ in Figure 2 (ignore number pairs on the edges). Observe that $\mathcal{P}(G)$ has two cycles: one between $A$ and $B$ and the other (self-loop) over $D$. Since they are vertex-disjoint, $G$ is strictly linear-recursive. Figure 9 (right) shows the production graph $\mathcal{P}(G')$ for the grammar $G'$ in Figure 8. Since $\mathcal{P}(G')$ has two self-loops that share $S$, $G'$ is linear-recursive but not strictly linear-recursive.*

It is possible to test in polynomial time if a given grammar $G$ is strictly linear-recursive. The algorithm starts by building the production graph $\mathcal{P}(G)$, then according to Definition 14, checks if any two cycles in $\mathcal{P}(G)$ share a vertex.

The main result of this paper is to show that dynamic, yet compact labeling is feasible for strictly linear-recursive grammars with any safe dependency assignment.

THEOREM 3. *Given any strictly linear-recursive workflow grammar $G$, for any safe dependency assignment $\lambda$, there is a compact dynamic labeling scheme for $G^\lambda$.*

The following section describes our labeling scheme.

## 4. VIEW-ADAPTIVE DYNAMIC LABELING

This section presents a compact view-adaptive dynamic labeling scheme for strictly linear-recursive workflows with safe views. The rationale behind our label design is explained as follows. Both data labels and view labels encode only partial (but orthogonal) reachability information. More precisely, a data label encodes only a subsequence of the run derivation that creates this data item, while a view label encodes only the fine-grained dependencies that are defined in this view. However, a combination of two data labels and a view label provides the complete information to infer the reachability between the two data items over this view.

We start with a preprocessing step in Section 4.1. Two independent tasks for labeling dynamic runs and labeling safe views are described in Sections 4.2 and 4.3, respectively. Section 4.4 presents how to efficiently answer queries using a combination of data labels and view labels. Finally, Section 4.5 analyzes the quality of our labeling scheme.

### 4.1 Preprocessing

As a preprocessing step, we assign a pair of numbers to each edge in the production graph. These pairs serve as unique ids for the edges, and will be used later to label runs and views. Let $G = (\Sigma, \Delta, S, P)$ be a strictly linear-recursive grammar and $\mathcal{P}(G)$ be its production graph. First of all, we fix an arbitrary ordering among the productions in $P$, and for each production $M \to W$, fix an arbitrary topological ordering among the modules in $W$. Let $p_k = M \to W$ be the $k$th production in $P$, and $M_i$ be the $i$th module in $W$, then we assign the edge from $M$ to $M_i$ in $\mathcal{P}(G)$ a pair $(k, i)$. Hereafter, we simply refer to this edge as $(k, i)$. In addition, we also fix an arbitrary ordering among all the (vertex-disjoint) cycles in $\mathcal{P}(G)$, and for each cycle, fix an arbitrary edge as the first edge of the cycle. We denote by $\mathcal{C}(s)$ the $s$th cycle in $\mathcal{P}(G)$ containing a list of number pairs.

EXAMPLE 11. *For the grammar $G$ in Figure 2, the pairs of numbers assigned to the edges in $\mathcal{P}(G)$ are shown in Figure 9. Note that the productions $p_1, p_2, \ldots, p_8$ are simply sorted by their subscripts. In Figure 2, all the modules in $W_1$ are sorted topologically as $a \to b \to A \to C \to c \to d$. Therefore, the edge from $S$ to $c$ in Figure 9 is assigned $(1, 5)$ because $p_1 = S \to W_1$ is the first production, and $c$ is the fifth module in $W_1$. Moreover, the two cycles in $\mathcal{P}(G)$ are denoted by $\mathcal{C}(1) = \{(2, 2), (4, 2)\}$ and $\mathcal{C}(2) = \{(6, 2)\}$.*

### 4.2 Labeling Dynamic Runs

Given a derivation of a run $R \in L(G)$, our goal is to assign a data label $\phi_r(d)$ to each data item $d$ in $R$ as soon as it is produced. The labeling is based on a tree representation for runs, called the *compressed parse tree*. In contrast to the traditional parse tree used for context-free grammars whose depth may be proportional to the size of the run, the depth of a compressed parse tree is always bounded by the size of the specification. We will see later that this property is critical to enable compact (logarithmic-size) data labels.

*Definition 15.* **(Compressed Parse Tree)** The *compressed parse tree* for a run $R$ is an ordered tree $\mathcal{T}(R)$, where each leaf node denotes an atomic module, and each non-leaf node denotes either a composite module (called the *composite node*), or a linear recursion (called the *recursive node*). The children of a composite node denote all the modules of a simple workflow produced by a production, and are ordered by a fixed topological ordering; and the children of a recursive node denote a sequence of nested composite modules obtained by unfolding a cycle in the production graph.

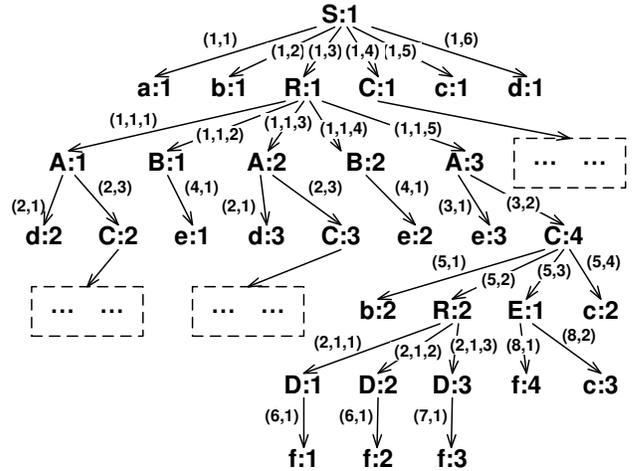

**Figure 10: Compressed Parse Tree**

EXAMPLE 12. *The compressed parse tree $\mathcal{T}(R)$ for the run $R$ in Figures 3 and 4 is shown in Figure 10 (ignore the edge labels), where $R\!:\!1$ and $R\!:\!2$ are recursive nodes. Note that $A\!:\!1, B\!:\!1, A\!:\!2, B\!:\!2, A\!:\!3$ (children of $R\!:\!1$) are obtained by unfolding the cycle between $A$ and $B$ in Figure 9. In a standard parse tree, they would be connected in a path.*

LEMMA 2. *Given a strictly linear-recursive workflow grammar $G$, for any derivation of a run $R \in L(G)$, the depth of the compressed parse tree $\mathcal{T}(R)$ is no greater than $2 * |\Delta|$, where $|\Delta|$ is the number of composite modules in $G$.*

We now describe the dynamic labeling algorithm. Given a derivation of a run $R$, we build $\mathcal{T}(R)$ in a top-down manner. During this process, we label each new edge and use the edge labels to construct labels for new data items on-the-fly. We next explain the design of data labels step by step.

Firstly, we describe the label for an edge in $\mathcal{T}(R)$. Let $e$ be an edge from $u$ to $v$ in $\mathcal{T}(R)$. We denote by $\phi_r(e)$ the label of $e$. (1) If $u$ is a composite node, then $e$ can be mapped to an edge $e'$ in $\mathcal{P}(G)$. Recall from Section 4.1 that each edge in $\mathcal{P}(G)$ is uniquely identified by a pair of numbers. Let $e' = (k, i)$, then $\phi_r(e) = (k, i)$; and (2) otherwise (if $u$ is a recursive node), let $u$ denote the $s$th cycle in $\mathcal{P}(G)$ starting from the $t$th edge. This can be determined by the first child of $u$. Let $v$ be the $i$th child of $u$, then $\phi_r(e) = (s, t, i)$.

Secondly, we use a sequence of edge labels to construct the label for an input port $i$ in $R$. We denote by $\phi_r(i)$ the label of $i$. Suppose $i$ is first created as the $x$th input port of a module $M$ during the derivation of $R$, and $M$ is denoted by a node $v$ in $\mathcal{T}(R)$. Let $e_1, e_2, \ldots, e_l$ be the path from the root node to $v$ in $\mathcal{T}(R)$, then $\phi_r(i) = \{\phi_r(e_1), \phi_r(e_2), \ldots, \phi_r(e_l), x\}$. For an output port $o$, $\phi_r(o)$ is defined similarly.

Finally, we use a pair of input and output port labels to construct the label for a data item (data edge) $d = (o, i)$ in $R$. We denote by $\phi_r(d)$ the label of $d$, then $\phi_r(d) = (\phi_r(o), \phi_r(i))$. Since $o$ and $i$ must be created by the same production, $\phi_r(o)$ and $\phi_r(i)$ differ only in the last one or two edge labels. The size of $\phi_r(d)$ can be reduced almost by half by factoring out the common prefix of $\phi_r(o)$ and $\phi_r(i)$.

EXAMPLE 13. *The edge labels for the compressed parse tree $\mathcal{T}(R)$ are shown in Figure 10. E.g., the edge from $R$:1 to $A$:3 is labeled by $(1, 1, 5)$, because $R$:1 denotes the first cycle in the production graph starting from the first edge (see Example 11), and $A$:3 is the fifth child of $R$:1. Next, we label the data items. E.g., consider $d_{21} = (o, i)$ in Figure 4, where $o$ is the first output port of $b$:2, and $i$ is first created as the second input port of $D$:1 (note that $i$ is also the second input port of $f$:1). Then, $\phi_r(d_{21}) = (\phi_r(o), \phi_r(i))$, where*

$$\phi_r(o) = \{(1,3), (1,1,5), (3,2), (5,1), 1\}$$
$$\phi_r(i) = \{(1,3), (1,1,5), (3,2), (5,2), (2,1,1), 2\}$$

### 4.3 Labeling Safe Views

Given a safe view $U = (\Delta, \lambda)$ over $G$, our goal is to create a view label $\phi_v(U)$ which can be combined with above data labels to infer reachability over $U$. Using the algorithm in Section 3.1, we first compute the full dependency assignment $\lambda^*$ by extending $\lambda$ to all the composite modules in $\Delta$.

Next, we define three functions, $\mathcal{I}$, $\mathcal{O}$ and $\mathcal{Z}$. Recall from Section 2.2 that $G_\Delta$ denotes the grammar obtained by restricting $G$ to $\Delta$. Let $\mathcal{P}(G_\Delta)$ be the production graph of $G_\Delta$, then $\mathcal{P}(G_\Delta)$ is a subgraph of $\mathcal{P}(G)$. Recall from Section 4.1 that each edge in $\mathcal{P}(G)$ is uniquely identified by a pair of numbers $(k, i)$. The input of $\mathcal{I}$ and $\mathcal{O}$ is an edge in $\mathcal{P}(G_\Delta)$, denoted by a pair $(k, i)$. The input of $\mathcal{Z}$ is a pair of edges in $\mathcal{P}(G_\Delta)$ of form $(k, i)$ and $(k, j)$. For simplicity, we also denote them by a triple $(k, i, j)$. The output of all three functions is a reachability matrix, which is defined next.

**Functions $\mathcal{I}$ and $\mathcal{O}$.** Given an edge $(k, i)$ in $\mathcal{P}(G_\Delta)$, let $p_k = M \to W$ be the $k$th production in $P$, and $M_i$ be the $i$th module in $W$, then (1) $\mathcal{I}(k, i)$ is defined as a reachability matrix from the inputs of $M$ to the inputs of $M_i$ (w.r.t. $\lambda^*$); and (2) $\mathcal{O}(k, i)$ is defined as a *(reversed) reachability matrix* from the outputs of $M$ to the outputs of $M_i$ (w.r.t. $\lambda^*$).

**Function $\mathcal{Z}$.** Given a pair of edges $(k, i)$ and $(k, j)$ in $\mathcal{P}(G_\Delta)$, let $p_k = M \to W$ be the $k$th production in $P$, and $M_i$ and $M_j$ be the $i$th and $j$th module in $W$, respectively, then $\mathcal{Z}(k, i, j)$ is defined as a *reachability matrix* from the outputs of $M_i$ to the inputs of $M_j$ (w.r.t. $\lambda^*$). Note that $\mathcal{Z}(k, i, j)$ is an empty matrix (with only false values) if $i \geq j$, since $M_i$ and $M_j$ are sorted in topological ordering.

Finally, $\phi_v(U)$ consists of all the above three functions, along with $\lambda^*(S)$ for the start module $S$. That is,

$$\phi_v(U) = \{\lambda^*(S), \mathcal{I}, \mathcal{O}, \mathcal{Z}\}$$

Basically, the above view label encodes all the fine-grained dependency information that is specific to this view and is necessary for our decoding algorithm given in Section 4.4.

EXAMPLE 14. *For the running example, we first label the default view $U_1 = (\Delta, \lambda)$ for which $\lambda^*$ is computed in Example 9, and is shown on the top of Figure 7. Using $\lambda^*$, we can compute the functions $\mathcal{I}$, $\mathcal{O}$ and $\mathcal{Z}$. E.g., consider the edge $(1, 5)$ from $S$ to $c$ in Figure 9. The first production $p_1 = S \to W_1$ is shown in Figure 2. $\mathcal{I}(1, 5)$ denotes the reachability from the inputs of $S$ (i.e., the initial inputs of $W_1$) to the inputs of $c$ (i.e., the fifth module in $W_1$); similarly, $\mathcal{O}(1, 2)$ denotes the (reversed) reachability from the outputs of $S$ (i.e., the final outputs of $W_1$) to the outputs of $b$ (i.e., the second module in $W_1$); and $\mathcal{Z}(1, 2, 5)$ denotes the reachability from the outputs of $b$ to the inputs of $c$ in $W_1$.*

$$\mathcal{I}(1,5) = \begin{bmatrix} 1 & 1 \\ 0 & 0 \end{bmatrix} \mathcal{O}(1,2) = \begin{bmatrix} 0 & 0 \\ 1 & 0 \\ 0 & 1 \end{bmatrix} \mathcal{Z}(1,2,5) = \begin{bmatrix} 0 & 0 \\ 0 & 0 \end{bmatrix}$$

*Similarly, we can label the other view $U_2 = (\Delta', \lambda')$ defined in Example 7, whose full dependency assignment is shown on the bottom of Figure 7. Using Figure 5, we have*

$$\mathcal{I}(1,5) = \begin{bmatrix} 1 & 1 \\ 0 & 1 \end{bmatrix} \mathcal{O}(1,2) = \begin{bmatrix} 1 & 0 \\ 1 & 1 \\ 1 & 1 \end{bmatrix} \mathcal{Z}(1,2,5) = \begin{bmatrix} 0 & 1 \\ 0 & 0 \end{bmatrix}$$

*As we can see above, the functions encoded by the view labels $\phi_v(U_1)$ and $\phi_v(U_2)$ may evaluate to different values for the same input. Moreover, they are defined over different domains. E.g., $\mathcal{I}(5, 1)$ is defined for $U_1$ but not for $U_2$.*

**Space-Efficient View Labeling.** By default, we precompute all the reachability matrices for $\mathcal{I}$, $\mathcal{O}$ and $\mathcal{Z}$, and materialize them in the view label. Alternatively, one can compute them on-the-fly by performing a graph search over the view of a specification during the query time. In general, more sophisticated approaches (e.g., [15, 24, 22]) can be used to label the view, in order to find a better balance between the overhead of labeling views and query efficiency. We will further explore this tradeoff in the experiments.

### 4.4 Decoding Data Labels with View Labels

Using only two data labels $\phi_r(d_1)$ and $\phi_r(d_2)$ and a view label $\phi_v(U)$, one can decide if $d_2$ depends on $d_1$ w.r.t. $U$ by a decoding predicate $\pi$. We first define in Section 4.4.1 two procedures used by $\pi$, namely Inputs and Outputs, and then describe $\pi$ in Section 4.4.2. Section 4.4.3 presents fast matrix multiplication used to achieve constant query time.

#### 4.4.1 Precedures Inputs and Outputs

Let $e$ be an edge from $u$ to $v$ in the compressed parse tree $\mathcal{T}(R)$. Given the edge label $\phi_r(e)$ (defined in Section 4.2) and a view label $\phi_v(U)$, our procedure Inputs computes a reachability matrix Inputs($\phi_r(e), \phi_v(U)$) by Algorithm 1.

**Case 1.** [Line 1 to Line 2] If $\phi_r(e) = (k, i)$, that is, if $u$ is a composite node, then Inputs computes a reachability matrix from the inputs of the module denoted by $u$ to the inputs of the module denoted by $v$, simply given by $\mathcal{I}(k, i)$.

**Case 2.** [Line 3 to Line 8] If $\phi_r(e) = (s, t, i)$, that is, if $u$ is a recursive node, then $v$ is the $i$th child of $u$. Let $M_1$, $M_2, \ldots, M_i$ be the modules denoted by the first $i$ children of $u$. They are a sequence of nested composite modules in $R_U$ obtained by unfolding the $s$th cycle in $\mathcal{P}(G)$ starting from the $t$th edge. `Inputs` finally computes a reachability matrix from the inputs of $M_1$ to the inputs of $M_i$ in $R_U$ by multiplying all $i - 1$ intermediate reachability matrices.

---

**Algorithm 1** *Procedure* `Inputs`

**Input:** $\phi_r(e) = (k, i)$ or $(s, t, i)$
       $\phi_v(U) = \{\lambda^*(S), \mathcal{I}, \mathcal{O}, \mathcal{Z}\}$
**Output:** `Inputs`$(\phi_r(e), \phi_v(U))$
1: **if** $\phi_r(e) = (k, i)$ **then**
2:   **return** $\mathcal{I}(k, i)$
3: **else** $\{\phi_r(e) = (s, t, i)\}$
4:   let $\mathcal{C}(s) = \{(k_1, i_1), (k_2, i_2), \ldots, (k_l, i_l)\}$
5:   // $\mathcal{C}(s)$ denotes the $s$th cycle in $\mathcal{P}(G)$ of length $l$
6:   let $\forall a \geq 1$, $k_{a+l} = k_a$ and $i_{a+l} = i_a$
7:   **return** $\prod_{a=1}^{i-1} \mathcal{I}(k_{t+a-1}, i_{t+a-1})$
8: **end if**

---

The other procedure `Outputs` is defined similarly, which computes a (reversed) reachability matrix for output ports.

EXAMPLE 15. *Let $e$ be the edge from $R:1$ to $A:3$ in Figure 10 and $U_1$ be the default view. $\phi_r(e) = (1, 1, 5)$ and $\phi_v(U_1)$ are explained in Examples 13 and 14. For this pair of labels, Algorithm 1 computes the reachability matrix from the inputs of $A:1$ to the inputs of $A:3$ in $R_{U_1}$. By Example 11, the first cycle is $\mathcal{C}(1) = \{(2, 2), (4, 2)\}$. Therefore,*

$$\text{Inputs}(\phi_r(e), \phi_v(U_1)) = \mathcal{I}(2,2) \times \mathcal{I}(4,2) \times \mathcal{I}(2,2) \times \mathcal{I}(4,2)$$

### 4.4.2 Decoding Predicate

Given a pair of data labels $\phi_r(d_1)$ and $\phi_r(d_2)$ and a view label $\phi_v(U) = \{\lambda^*(S), \mathcal{I}, \mathcal{O}, \mathcal{Z}\}$, our goal is to evaluate $\pi$ to true iff $d_2$ depends on $d_1$ w.r.t. $U$. Due to space constraints, we sketch only the main cases, where both $d_1$ and $d_2$ are intermediate data items of $R$. The complete description can be found in the full version of this paper [6]. Let $\phi_r(d_1) = (\phi_r(o_1), \phi_r(i_1))$ and $\phi_r(d_2) = (\phi_r(o_2), \phi_r(i_2))$, then $d_2$ depends on $d_1$ w.r.t. $U$ iff $i_2$ is reachable from $o_1$ in $R_U$. Let $\phi_r(o_1) = \{l_1, x\}$ and $\phi_r(i_2) = \{l_2, y\}$, where $l_1$ and $l_2$ are two lists of edge labels. Suppose during the derivation of $R$, $o_1$ is first created as the $x$th output port of some module $M_1$ and $i_2$ is first created as the $y$th input port of some module $M_2$. Suppose $M_1$ and $M_2$ are denoted by two nodes $v_1$ and $v_2$ in the compressed parse tree $\mathcal{T}(R)$.

**Case 1.** If $l_1 = l_2$ or one is a prefix of the other, that is, $v_1 = v_2$ or one is an ancestor of the other in $\mathcal{T}(R)$, then $M_1 = M_2$ or one is derived from the other. Thus, $i_2$ is not reachable from $o_1$ in $R_U$, and $\pi$ evaluates to false.

**Case 2.** Otherwise, suppose $l_1$ and $l_2$ agree on the first $l-1$ edge labels, but differ on the $l$th edge label. Moreover, let the length of $l_1$ and $l_2$ be $p$ and $q$, respectively. That is,

$$l_1 = \{\phi_r(e_1), \ldots, \phi_r(e_{l-1}), \phi_r(e_l), \ldots, \phi_r(e_p)\}$$
$$l_2 = \{\phi_r(e_1), \ldots, \phi_r(e_{l-1}), \phi_r(e'_l), \ldots, \phi_r(e'_q)\}$$

where $\phi_r(e_l) \neq \phi_r(e'_l)$. We denote by $v = LCA(v_1, v_2)$ the *least common ancestor* of $v_1$ and $v_2$ in $\mathcal{T}(R)$. Let $e_l$ be an edge from $v$ to $v'_1$ and $e'_l$ be an edge from $v$ to $v'_2$. Let $M'_1$ and $M'_2$ be the module denoted by $v'_1$ and $v'_2$, respectively.

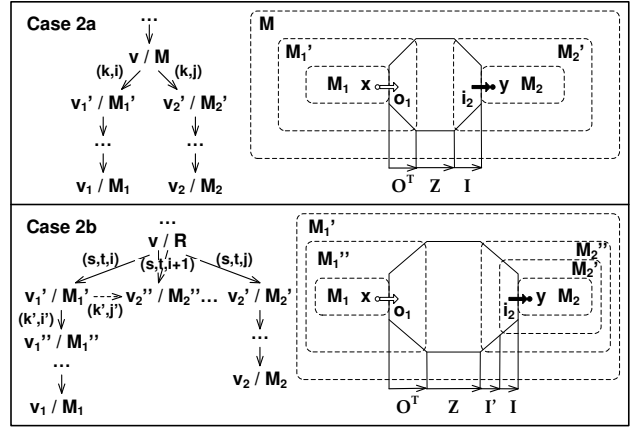

**Figure 11: Main Cases of Decoding Predicate**

**Case 2a.** If $\phi_r(e_l) = (k, i)$ and $\phi_r(e'_l) = (k, j)$, that is, $v = LCA(v_1, v_2)$ is not a recursive node, then we compute

$$O = \Pi_{a=l+1}^{p} \text{Outputs}(\phi_r(e_a), \phi_v(U))$$
$$I = \Pi_{a=l+1}^{q} \text{Inputs}(\phi_r(e'_a), \phi_v(U))$$

and $Z = \mathcal{Z}(k, i, j)$. As illustrated by the top right corner of Figure 11, $O$ is the (reversed) reachability matrix from the outputs of $M'_1$ to the outputs of $M_1$, $Z$ is the reachability matrix from the outputs of $M'_1$ to the inputs of $M'_2$, and $I$ is the reachability matrix from the inputs of $M'_2$ to the inputs of $M_2$. Thus, $O^T \times Z \times I$ gives the reachability matrix from the outputs of $M_1$ to the inputs of $M_2$, where $O^T$ denotes the transpose of $O$. So $\pi$ evaluates to $(O^T \times Z \times I)[x, y]$.

**Case 2b.** If $\phi_r(e_l) = (s, t, i)$ and $\phi_r(e'_l) = (s, t, j)$, that is, $v = LCA(v_1, v_2)$ is a recursive node, we consider the case where $i < j$. The other case where $i > j$ can be handled in a similar manner. First of all, if $p = l$, that is, $v_1 = v'_1$ and $M_1 = M'_1$, then $M_2$ is derived from $M_1$. By Case 1, we know that $i_2$ is not reachable from $o_1$ in $R_U$. So $\pi$ evaluates to false. Otherwise, as illustrated by the bottom right corner of Figure 11, using a similar decoding process to Case 2a, $\pi$ evaluates to $(O^T \times Z \times I' \times I)[x, y]$ (see [6] for details).

### 4.4.3 Fast Matrix Multiplication

To achieve constant query time, we need to show that `Inputs` and `Outputs` can be implemented in constant time.

LEMMA 3. *Given a fixed strictly linear-recursive grammar $G$, for any edge label $\phi_r(e)$ and any data label $\phi_v(U)$, `Inputs` and `Outputs` can be computed in constant time.*

PROOF. Consider Case 2 in Algorithm 1. First observe the repeated pattern of length $l$ in the $i - 1$ intermediate reachability matrices. Let $X$ be the multiplication of the first $l$ matrices. So we only need to efficiently compute $X^{\lfloor i-1/l \rfloor}$. Further observe the repeated pattern in the sequence $X, X^2, \ldots, X^{\lfloor i-1/l \rfloor}$. Suppose any module has at most $c$ input or output ports. Note that $c$ is a constant for a fixed $G$. Since each matrix has at most $2^{c \times c}$ possible boolean values, we can find in constant time $a$ and $b$ such that $a < b <= 2^{c \times c} + 1$ and $X^a = X^b$. Once $a$ and $b$ are found, $X^{\lfloor i-1/l \rfloor}$ can be computed in constant time. □

**Query-Efficient View Labeling.** To speed up the query processing, one can also pre-compute $a$ and $b$ for each recursion in the view, and materialize $a$ and $b$ (as well as $X^1, X^2, \ldots, X^b$) in the view label. In contrast to space-efficient view labeling (Section 4.3), this is the other extreme alternative that will be compared in the experiments.

## 4.5 Labeling Scheme Quality Analysis

We analyze the *label length* and *construction time* for both data labels and view labels, as well as the *query time* for comparing a pair of data labels and a view label. Note that we take the size of a specification as constant [4, 5], and measure the complexity in terms of the size of the run. We next show that all the above parameters, guaranteed by our labeling scheme, are *optimal* up to a constant factor.

THEOREM 4. *Let $(\phi_r, \phi_v, \pi)$ be our view-adaptive dynamic labeling scheme for a strictly linear-recursive specification $G$.*

1. *logarithmic label length and linear total construction time for data labels: for any derivation of a run $R \in L(G)$ with $n$ data items and for any data item $d$ in $R$, $\phi_r(d)$ has $O(\log n)$ bits, and all data labels can be constructed dynamically in a total of $O(n)$ time.*

2. *constant label length and constant construction time for view labels: for any safe view $U$ over $G$, $\phi_v(U)$ has $O(1)$ bits and can be constructed in $O(1)$ time.*

3. *constant query time: for any pair of data labels $\phi_r(d_1)$ and $\phi_r(d_2)$ and for any view label $\phi_v(U)$, $\pi(\phi_r(d_1), \phi_r(d_2), \phi_v(U))$ can be evaluated in $O(1)$ time.*

PROOF. (Sketch) Lemma 2 ensures $O(\log n)$ data label length. Lemmas 2 and 3 ensure $O(1)$ query time. □

**User-Defined Views.** Our view-adaptive labeling scheme can be extended to handling more general types of views, where users may create their own composite modules (rather than using *pre-defined* ones) or may hide ports or data edges. Details can be found in the full version of this paper [6].

## 5. EXPERIMENTAL EVALUATION

We now empirically evaluate the effectiveness of our view-adaptive labeling approach. Section 5.2 reports the main cost of labeling, which is labeling runs. Section 5.3 explores the tradeoff between the overhead of labeling views and query time by comparing three alternative implementations. Section 5.4 demonstrates the superiority of view-adaptive labeling over the state-of-the-art technique [5] when applied to label multiple views. Section 5.5 identifies important factors that influence the performance of view-adaptive labeling.

## 5.1 Experimental Setup

**Real-Life and Synthetic Datasets.** Our real-life scientific workflows were collected form the *myExperiment* workflow repository [19]. We observed that almost all of them have fairly simple recursive patterns. For simplicity, we report only the results for one representative workflow, called BioAID. It is denoted by a strictly linear-recursive grammar with 112 modules (16 are composite) and 23 productions (7 are recursive). Each production produces a simple workflow with at most 19 modules, and each module has at most 4 input ports and 7 output ports. In Section 5.5, we also evaluate a family of synthetic workflows. Due to the absence of real workflow executions, we simulated runs by applying a random sequence of productions, varying their sizes (i.e., the number of data items) from $1K$ to $32K$ by a factor of 2. The derivations of runs were recorded and used as dynamic inputs to labeling schemes. In addition, we obtained safe views by enumerating all possible proper subsets of composite modules and assigning random input-output dependencies to atomic modules. All the data are stored as XML files whose parsing time is omitted from the results.

**Labeling Schemes.** Our view-adaptive dynamic labeling scheme is denoted by *FVL* for (F)ine-grained (V)iew-adaptive (L)abeling. We implemented three variants: (1) *Default FVL* (Section 4.3) (2) *Space-Efficient FVL* (Section 4.3) and (3) *Query-Efficient FVL* (Section 4.4.3). They use the same dynamic algorithm to label runs, but differ in how views are labeled, which affects query efficiency. We also compared *FVL* with the state-of-the-art scheme, called *DRL* [5], for (L)abeling (D)ynamic runs of (R)ecursive workflows. All the labeling schemes were implemented in Java.

**Evaluation Methodology.** To evaluate labeling overhead, we measure both label length (space overhead) and construction time (time overhead) for data labels and view labels, respectively. For data labels, each data point in the result is an average over 100 sample runs. We also measure the query time. Each data point for query time is an average over $10^6$ sample queries. All the experiments were performed on a local PC with Intel(R) Core(TM) i7-2600 3.40GHz CPU and 4GB memory running Windows 7 Professional.

## 5.2 Overhead of Labeling Runs

We first evaluate the overhead of labeling runs using *FVL* and *DRL*. Note that *FVL* is *view-adaptive*: the data labels created for one run can be re-used to answer queries over all safe views. In contrast, *DRL* is not view-adaptive: a run must be re-labeled for each view. Here, the comparison between them focuses on the case where only one default view is defined over the workflow. A more meaningful comparison for multiple views will be carried out in Section 5.4.

Figure 12 reports the maximum and average length of data labels created by *FVL* and *DRL*. We denote them by *FVL-max*, *FVL-avg*, *DRL-max* and *DRL-avg*, respectively. A careful analysis of Figure 12 can show that all four lines are nearly parallel to the asymptotic line $f(x) = \log x$. This implies that both *FVL* and *DRL* produce compact data labels of logarithmic length with a constant factor close to 1. Surprisingly, *FVL-avg* (*FVL-max*) is even shorter than *DRL-avg* (*DRL-max*) by about 5 bits. This small improvement is due to the compact design of data labels in *FVL* which encode only the structure of runs.

Figure 13 reports the construction time of data labels for *FVL* and *DRL*. While both build all data labels in linear time, *FVL* is faster than *DRL* by about 10% for large runs.

## 5.3 View Labeling Cost vs. Query Efficiency

Next, we evaluate the overhead of labeling views as well as the query time, and explore the tradeoff between them by comparing three variants of *FVL*: (1) *Default FVL* pre-computes all reachability matrices for the three functions $\mathcal{I}$, $\mathcal{O}$ and $\mathcal{Z}$, and materializes them in the view label (Section 4.3); (2) *Space-Efficient FVL* pre-computes only the full dependency assignment for each view, and thus any access to $\mathcal{I}$, $\mathcal{O}$ and $\mathcal{Z}$ will be answered by performing a graph search over the view of a specification at query time (Section 4.3); and (3) *Query-Efficient FVL* materializes, in addition to $\mathcal{I}$, $\mathcal{O}$ and $\mathcal{Z}$, all intermediate states of fast matrix multiplication for each recursion in the view (Section 4.4.3).

In the experiments, we label three safe views, namely, small view, medium view and large view, with varying sizes and random dependency assignments. We estimate the size of a view by the number of composite modules that can expand. The three views contain 2, 8 and 16 composite modules, respectively. Figure 14 shows the length of view labels created by all three variants of *FVL*. As expected, *Query-Efficient FVL* creates the longest labels for all three views.

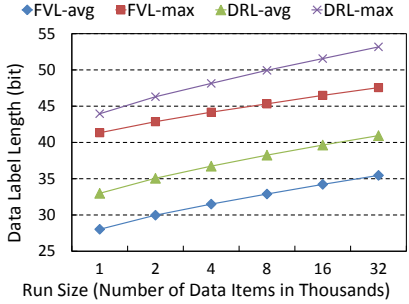
Figure 12: Space Overhead (run)

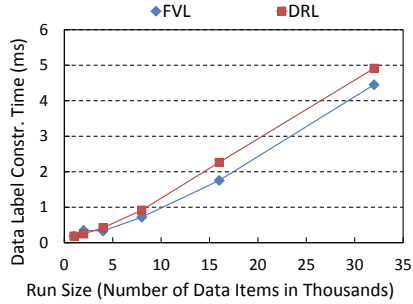
Figure 13: Time Overhead (run)

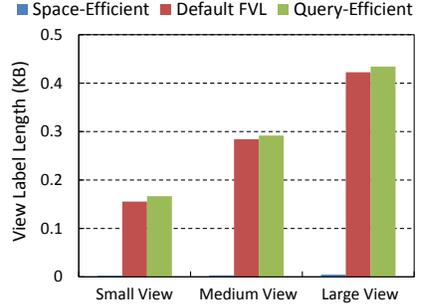
Figure 14: Space Overhead (view)

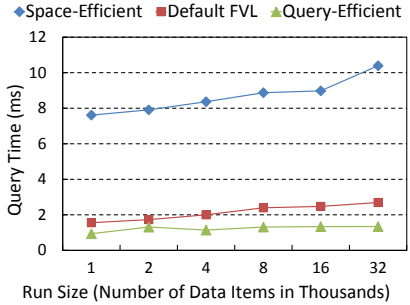
Figure 15: Query Time

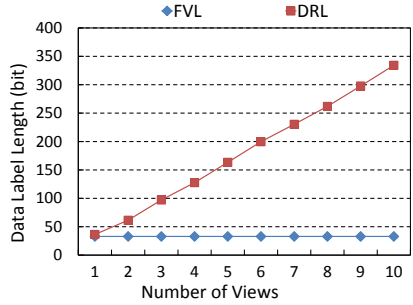
Figure 16: FVL vs DRL (space)

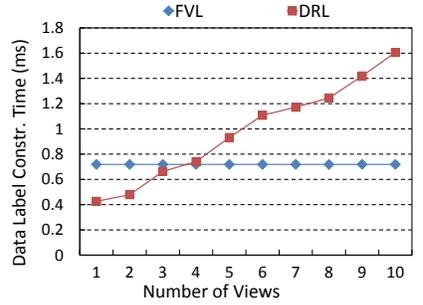
Figure 17: FVL vs DRL (time)

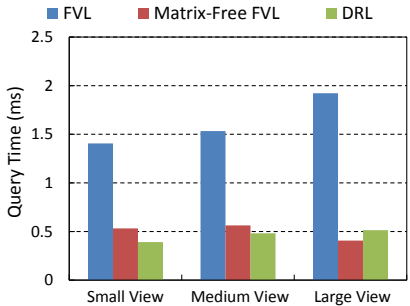
Figure 18: FVL vs DRL (query)

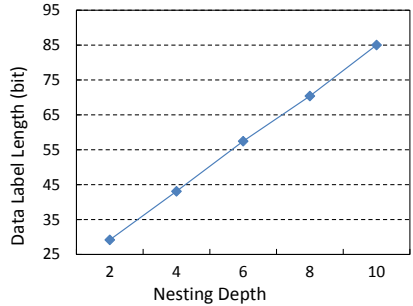
Figure 19: Nesting Depth

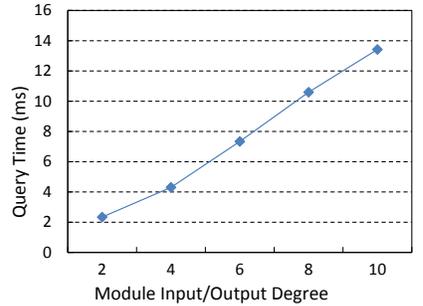
Figure 20: Module Degree

However, compared with *Default FVL*, the extra space overhead is small (less than 8 bytes), since views typically have a small amount of recursions. On the other hand, *Space-Efficient FVL* creates almost no index for each view (less than 5 bytes). The results for construction time, not shown, reveal a similar trend. While *Query-Efficient FVL* labels the large view in 0.62 *ms*, *Space-Efficient FVL* needs only 0.08 *ms*. Comparing Figures 12 and 14 also shows that the main overhead of *FVL* lies in the labeling of runs, e.g., the data labels for a small run with 1*K* data items take a total of 5*KB*, while the view label created by *Query-Efficient FVL* for the large view takes only 0.4*KB*. The overall difference is even bigger, since for a given workflow, the number of runs is typically much greater than the number of views.

After runs and views are both labeled (independently), we generate sample queries by randomly selecting two data items in the same run (with varying size) and randomly selecting one out of the three views. The query time for the three variants of *FVL* is reported in Figure 15. Compared to Figure 14, we can see a clear tradeoff between the overhead of labeling views and query efficiency. *Query-Efficient FVL* and *Default FVL* are faster than *Space-Efficient FVL* by almost one order of magnitude. *Query-Efficient FVL* is also significantly faster than *Default FVL* (by about 40% for large runs), while as shown in Figure 14, it takes only small extra space overhead (less than 2% for the large view).

Finally, we should notice that all three variants of *FVL* achieve constant view label length and constant query time, in terms of the size of the run. In other words, there is only a constant tradeoff between space and time for the three approaches. Therefore, *Query-Efficient FVL* is preferable to the other two variants, since it enables the fastest query processing with little extra labeling overhead. All the above results also validate our complexity analysis in Theorem 4.

## 5.4 Advantage of View-Adaptive Labeling

We now compare *FVL* against *DRL* when multiple views are defined over the same workflow. Since *DRL* applies only to the coarse-grained model with black-box dependencies, to make a meaningful comparison we randomly generate 10 medium-size views with black-box dependencies.

First, we compare the labeling overhead of *FVL* and *DRL*. Our focus is on the overhead of labeling runs, which is the main cost. We fix the size of runs to be 8*K* (data items), and vary the number of views from 1 to 10. Figure 16 shows the total length of data labels assigned to one data item. Since *FVL* is view-adaptive, the data label created for one data item can be re-used to query over multiple views. Therefore, in Figure 16, the total length for *FVL* remains constant. In contrast, given a data item, *DRL* has to maintain one data label for each view separately. So in Figure 16, the total length for *DRL* grows linearly with the number of views.

A similar result for the total construction time can be observed in Figure 17. Note that *DRL* is faster than *FVL* for one view, since *DRL* labels the medium-size view of a run, which is smaller than the original run. However, when there are more than 3 views, *FVL* is more time-efficient.

We next compare the query time of *FVL* and *DRL*. In order to achieve a fair comparison, we take the most *query-efficient* variant of both *FVL* and *DRL*. Since our comparison can only use coarse-grained views, many of the reachability matrices involved in the decoding of *FVL* are complete matrices (i.e., with only true values). So we also implemented a simplified version of *FVL*, called *Matrix-Free FVL*, which is optimized for coarse-grained views by avoiding redundant matrix multiplications in the decoding.

We evaluate the above three approaches over three coarse-grained views with varying sizes. As shown in Figure 18, *FVL* is about 4 times slower than *DRL*, but by removing redundant computations for coarse-grained views, *Matrix-Free FVL* achieves almost same query time as *DRL*.

## 5.5 Important Factors

Finally, we examine the effectiveness of *FVL* over a variety of synthetic workflows. The goal is to identify factors that affect *FVL*. In particular, we consider: (1) *workflow size*: the number of modules in a simple workflow (default = 40); (2) *module degree*: the number of input/output ports of a module (default = 4); (3) *nesting depth*: the depth of nested composite modules (default = 4); and (4) *recursion length*: the number of composite modules in a recursion (default = 2). We created a family of synthetic workflows by varying each of the four parameters and fixing the rest to be the default value. For each workflow, we evaluate (1) the overhead of labeling a run $R$ with $8K$ data items; (2) the overhead of labeling a safe view $U$ with all composite modules and random dependency assignment; and (3) the query time for data items in $R$ over $U$. Due to space constraints. we show only the results for two key factors that affect *FVL*.

One factor that has high impact on the data label length is nesting depth. As shown in Figure 19, the (average) data label length created by *FVL* grows linearly with the nesting depth, because the nesting depth determines the depth of the compressed parse tree which is used to build data labels.

Another factor that has high impact on the query time is module degree. As shown in Figure 20, the query time for *Query-Efficient FVL* grows almost linearly with the module degree. This is mainly because the module degree determines the cardinality of reachability matrices, and multiplying large matrices at query time can be expensive.

## 6. CONCLUSIONS

This paper considers the problem of efficiently answering reachability queries over views of workflow provenance graphs. For that we design a novel *view-adaptive* labeling scheme that supports *fine-grained dependencies* between inputs and outputs of modules and combines *static* labeling of views with *dynamic* labeling of data items. In particular, we identify a natural class of *safe views* over *strictly linear-recursive* workflows for which dynamic, yet *compact* labeling is feasible. The experimental results demonstrate the advantage of our view-adaptive labeling approach over the state-of-the-art technique [5] when applied to label multiple views. Previous work [12] considers efficient evaluation of XPath queries over XML views. Extending our work to similarly rich query constructs in the context of workflow views is an interesting direction for future research.

## 7. ACKNOWLEDGMENTS

We thank the anonymous reviewers for their helpful comments. This work was supported in part by the US National Science Foundation grants IIS-0803524 and IIS-0629846, by the European Research Council under the European Union's Seventh Framework Programme (FP7/2007-2013) / ERC grant 291071-MoDaS, by the Israel Ministry of Science, and by the Binational (US-Israel) Science Foundation.